\documentclass[14pt]{extarticle}
\usepackage{graphicx}
\usepackage{helvet}
\usepackage{setspace}
\usepackage{rotating}
\usepackage{booktabs}
\usepackage{lscape}
\usepackage[superscript]{cite}
\usepackage{longtable}
\usepackage{tikz} 
\def\checkmark{\tikz\fill[scale=0.4](0,.35) -- (.25,0) -- (1,.7) -- (.25,.15) -- cycle;}
\def\tinycheckmark{\tikz\fill[scale=0.2](0,.35) -- (.25,0) -- (1,.7) -- (.25,.15) -- cycle;}
\begin{document}
\thispagestyle{empty}
\fontfamily{phv}\selectfont{
{\huge \noindent How to estimate solid-electrolyte-\newline interphase features when screening electrolyte materials}
\\ \\
{\scriptsize Tamara Husch, Martin Korth$^{*}$\\
Institute for Theoretical Chemistry, Ulm University, Albert-Einstein-Allee 11, 89069 Ulm (Germany)\\
$^{*}$ Corresponding author email: martin.korth@uni-ulm.de\\ \\


\noindent {\bf \fontfamily{phv}\selectfont{Abstract}}\\

\noindent
Computational screening of battery electrolyte components is an extremely challenging task
because very complex features like solid-electrolyte-interphase (SEI) formation and graphite exfoliation
need to be taken into account at least at the final screening stage.
We present estimators for both SEI formation and graphite exfoliation
based on a combinatorial approach with quantum chemistry calculations on model system reactions,
which can be applied automatically for a large number of compounds
and thus allows for a systematic first assessment of the relevant properties within screening approaches.
Thermodynamic effects are assessed with quantum mechanical calculations,
a more heuristic approach is used to estimate kinetic effects.
\\ \\

\noindent {\bf \fontfamily{phv}\selectfont{Introduction}}\\

\noindent
Chemical energy storage is of high importance for meeting future energy demands in an eco-friendly way.\cite{review_all_general-motors_2010}
With substantial improvements in the field of advanced electrode materials,
the limitations of current electrolyte systems are more and more often found to be a roadblock for further improvements of battery technology:
\cite{review_all_scrosati_2011,review_all_tarascon_2010,review_all_aurbach_2011b,review_all_goodenough_2014,review_all_winter_2013}
With increasing chemical potential differences between improved anode and cathode materials,
especially the required electrochemical stability of the electrolyte is a weak point.
Research focusing on the integrated experimental and theoretical screening of electrolyte components
has given promising results, but is still rather at it's beginning,\cite{Korth2015}
unlike corresponding work on electrode materials.\cite{Jain2013}
High-throughput work on electrolyte materials has so far only considered comparably simple single-molecule properties which were treated with very basic DFT approaches,\cite{Cheng2015}
though computational chemistry offers better suited tools for this purpose.\cite{Korth2014}
We have recently shown how to efficiently estimate collective properties like viscosities and flash points,\cite{Husch2015,balducci}
which are of utmost importance for making reasonable suggestions for experimental investigations,
and how to efficiently perform large-scale screening of several million compounds.\cite{liair}
Here we turn to combinatorial quantum chemistry to do the next step,
which is including estimators for complex, reactivity-related estimators beyond single-molecule stability.

Current lithium-ion battery (LIB) technology most often relies on mixtures of
cyclic and linear carbonates like ethylene carbonate (EC) and dimethyl carbonate (DMC) as electrolyte solvents
combined with inorganic lithium compounds like hexafluorophosphate (LiPF$_6$) as salts.\cite{review_electrolytes_kang_2004}
Such electrolyte systems show reasonable performance with respect to all relevant properties,
like electrochemical stability windows, melting, boiling and flash points, dielectric constants, viscosity,
ionic and electronic conductivity, toxicity and price.\cite{review_electrolytes_kang_2011}
The electrochemical stability is a special case, because it is only achieved through a
passivating layer, the so-called solid-electrolyte-interphase (SEI),
which is build up during initial charge/discharge cycling mostly from the decomposition of the least stable electrolyte component, EC,
and prevents ongoing electrolyte decomposition.\cite{review_all_goodenough_2010}
Estimating the electrochemical stability of an electrolyte system is thus greatly complicated
by the fact that it is not a single-molecule feature but the result of the reaction
of the electrolyte components with each other at the electrode surface.
The formation of stable SEI films is of essential importance for the battery performance,
but is hard to characterize experimentally and cannot yet be predicted with computational models.\cite{wft_electrolytes_leung_2013,Korth2015}
Accordingly many somewhat different reports about the composition of SEI films exist,
though there seems to be a rather broad agreement on the main components of the most important systems (see below).\cite{review_sei_novak_2010,review_sei_xu_2010,review_sei_xu_2012}
On the theoretical side researchers face a dilemma: Accurate enough wave function theory (WFT) methods
can not yet be applied in dynamical simulation of SEI formation, while lower-level approaches like (reactive) force fields
or even density functional theory (DFT) methods were found to be too inprecise for conclusive dynamical studies.\cite{wft_electrolytes_leung_2013,Korth2015} 
Most existing computational screening studies thus resort to the estimation of electrochemical stability windows
based on single-molecule orbital energy values. 
We propose a feasible approach to predict the reaction products of SEI formation based on heuristic rules included in a combinatorial approach
followed by the application of various different existing tools from quantum chemistry, chemical engineering and chemoinformatics.
This method does not aim to elucidate the underlying mechanisms, but automatically predict
the main components of the SEI formed from an arbitrary compound or mixture in a systematic way.
The (inhomogeneous) spacial structure of SEI films can not be resolved, but estimating the main (and other possible) components is a valuable step also for the analysis of experimental work.
Approaches to predict reactivity have been published before,\cite{baldi,goodman}
but can not be directly transfered to our problem, because of the unusual mechanism and high complexity of electroorganic chemistry.\cite{oec1,oec2}
Another complex property that would be desirable to include in screening approaches is graphite exfoliation.
While the most common electrolyte solvent, EC, is able to form a stable SEI, the structurally closely related propylene carbonate (PC) leads to graphite exfoliation.
We propose an estimator for graphite exfoliation based on the interlayer distance of model graphite intercalation compounds (GICs),
as suggested by Tasaki/Winter,\cite{wft_electrolytes_tasaki-winter_2011,screening_dft_intercalation_tasaki_2014} but construct GICs automatically not only for the original solvent molecule but all predicted SEI compounds,
to capture the effect of the decomposition processes that are occurring before and during graphite exfoliation.
The inclusion of these complex properties should lead to more realistic theoretical suggestions for better electrolyte materials.
Furthermore, the results are valuable for assisting in the evaluation of experimental data and the setting up of in-depth theoretical investigations.
\\ \\

\noindent {\bf\fontfamily{phv}\selectfont{Methodological details}}\\

\noindent
Below we focus on the main features of our approach, further details can be found in the supplementary information.
Lithiation of organic compounds is possible automatically with an in-house program.
The energy of the structures is determined with PM6-DH+\cite{Korth2010} using MOPAC2012.\cite{mopac}
Our approach is not limited to semi-empirical methods, any electronic structure theory method could equally well be employed.
We carried out B3LYP  and BP86 DFT calculations with TURBOMOLE 6.4\cite{ahlrichs89,tm64}, using D2 dispersion corrections\cite{DFT-D2}, the RI approximation for two-electron integrals\cite{eichkorn95,eichkorn97}
and TZVP AO basis sets\cite{schaefer94} for comparison with our semi-empirical approach, details of which can be found in the supplementary information.
Energies and orbital eigenvalues are computed at SQM and B3LYP DFT level.
Solubilities of the products in the input solvent are calculated using COSMOtherm \cite{Klamt2011} on top of SQM or BP86 DFT calculation.
Tanimoto coefficients are determined with OpenBabel. \cite{OpenBabel}
The results presented here are thus based on simplified model systems and approximate computational methods
and should be taken with the appropriate care (as simpler problems were
shown to sometimes require much more advanced methods).\cite{methodstudy2}
\\ \\

\noindent {\bf \fontfamily{phv}\selectfont{Estimating critical complex properties: SEI composition}}\\

\noindent
An important goal for research into advanced electrolyte system
would be the identification of criteria for forming functional SEI films, based on the molecular composition of electrolyte systems.
On the experimental side, several combined studies of battery performance (an indicator for functional SEI film)
and SEI composition were published in the past (for an overview see \cite{review_sei_novak_2010}, \cite{Xu2014}),
but systematic studies on a wider range of compounds are still missing. 
On the theoretical side, the problem first of all lies with estimating SEI composition as the prerequisite
for correlating computational results on SEI components with experimental data on battery performance.
Most commonly, research into SEI formation is either based on 'static', higher-level studies on clusters of one or more solvent and salt molecules,
or 'dynamical', low- to mid-level studies on the dynamics of a periodic model system of solvent and salt molecules on a more or less realistic electrode surface.\cite{aimd_sei_leung_cathode-material_2012,Korth2015}
Our method is complementary to both approaches and aims at the identification of the major components of the SEI.
Therefore, a 'redox fingerprint' of the involved (arbitrary) molecules is constructed.
The denotation 'redox fingerprint' refers to all generated reactions for the specified input parameters (input molecule, number of reactants, number of transferred electrons).
All possible reactions for a specified number of reactants and electrons are generated based on combinatorial considerations and heuristic rules.
In a second step, the 'redox fingerprint analysis' (RFPA), the probability to encounter the reaction products in the SEI is evaluated
with estimators for their thermodynamic stability, solubility and kinetic barriers using the quantum chemistry and chemical engineering methods we have benchmarked in our previous studies.
Figure 1 shows a schematic overview of the RFPA. 
\begin{figure}[h!t]
\begin{center}
\includegraphics[width=\textwidth]{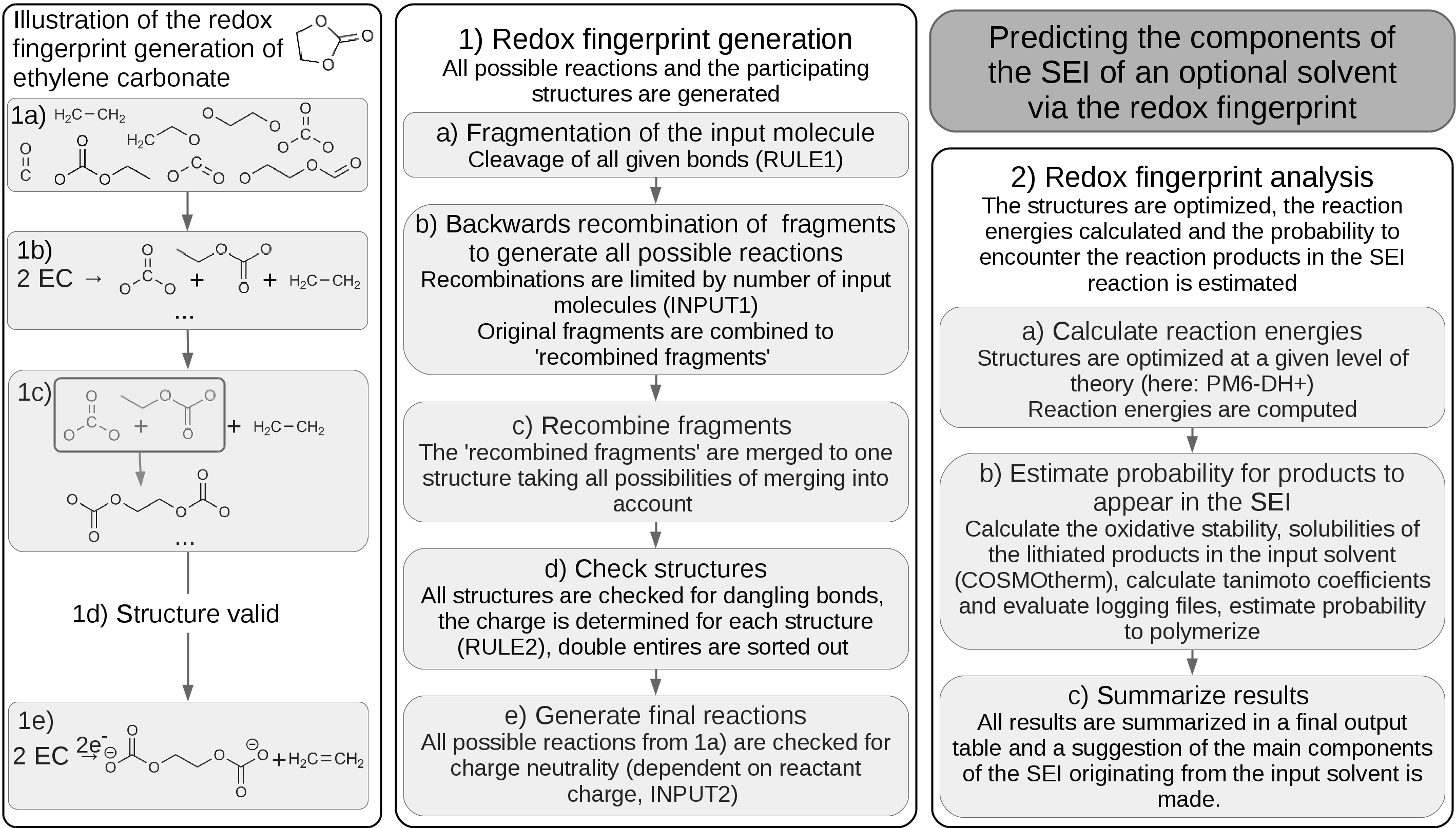}
\fontfamily{phv}\selectfont{\caption{Schematic overview of the redox fingerprint analysis (RFPA) approach with an exemplary illustration of the formation of the (experimentally identified) main component, ethylene dicarbonate, from ethylene carbonate to clarify the redox fingerprint generation.}}
\label{fig:fig1}
\end{center}
\end{figure}
The prediction of the main components of the SEI consists of two steps. In the first step the redox fingerprint of the input molecule is constructed. 
The generation of all possible reactions starts with the fragmentation of the input molecule (1a). All fragments are generated that can be obtained through the cleavage of defined bonds. 
The bonds that are allowed to be broken are identified with heuristic rules. These rules have been derived from textbook knowledge of likely bond cleavages. 
The rules currently comprise the cleavage of: Single bonds between a hetero atom with a coordination greater than one and an arbitrary heavy atom, single bonds between a cyanide group and an arbitrary heavy atom, 
single bonds in the neighborhood of carbonyl groups, single bonds between halogen atoms (coordination equals one) and an arbitrary heavy atom. 
The rules are not extensive and can be expanded easily. 
Note that though products of rearrangment reactions can occure, not all possible products resulting from for instance proton transfer are taken into account yet, which can be included through the introduction of additional
cleavage rules.
Currently the most common elements in organic chemistry are implemented: H, B, C, N, O, P, S, Si, F and Cl.
The resulting fragments are then combined backwards to generate all possible reactions, {\it i.e.} it is determined which (and how many) fragments can result from the fragmentation of a given maximum number of input molecules (1b). A valid reaction conserves all atoms of the input molecules. The number of maximum reactant molecules is an input parameter that specifies the maximum number of molecules taking part in one reaction. The default value is currently set to two molecules. It has to be limited because otherwise an infinitely high number of reactions would be possible. However, the limitation prohibits the formation of oligomers.
What we have called a recombination above is actually not already a specific chemical species, as there often is more than one possibility to combine a given number of molecular fragments into a new molecule (1b), 
though all these variants will similarly match to the above constructed decomposition reaction. All possibilities to recombine the original fragments to chemical structures are determined for each possible reaction . The corresponding structures are generated in the next step (1c). 
Not all generated structures will be reasonable products or unique. All unreasonable products and double entries are sorted out in the next step (1d).
(We do not use information on double entries etc. to assess a possible influence on the likeliness of the occurence of certain products.)
The charge of each structure is determined based on heuristic rules because the underlying mechanism is not further clarified. These rules are based on the valence of the atom in the structure. A negative charge is mostly attributed to more electronegative atoms and a positive charge to more electropositive atoms. However, we chose to not allow each charge situation, for example carbon is only supposed to be stable in the uncharged state. For each element for which no rules have been implemented yet, the valence has to correspond to the standard valence, but the rules can be easily extended. 
Not all combinations satisfy the requirement of balanced charge for each specific reaction (1e). The charge balance depends on the charge initially chosen for the input molecule, which is another input parameter of our approach. Conveniently, the procedure can easily be completed for different input charge values, which can model to some extend the decaying effect of the electrode polarization on the solvent molecules:
A high input charge will more likely reflect the situation near the electrode, while a lower input charge should be a more reasonable estimate on the electrolyte side of the SEI.
Our approach does not on the other hand generate information about the spatiotemporal inhomogeneities of these layers.
Accordingly the final reactions, the redox fingerprint, are generated. The most likely reduction processes identified in the literature are one and two electron processes. Therefore, we look at these processes by default.

In the next step, the generated reactions are examined. The reaction energies are calculated with quantum chemical methods to evaluate the thermodynamics of the reactions (2a). For the thermodynamically most favorable reactions, the probability of the reaction products to appear in the SEI is estimated. The estimation is based on heuristic rules because the kinetics of the reaction are not readily available (2b). The estimators include the solubility of the reaction products in the input solvent, the oxidative stability, the Tanimoto coefficient (similarity measure) and the number of needed fragmentations and recombinations to generate the reaction products. An empirical estimate of the polymerizability is included because the restriction to two molecules prohibits oligomerization.
Though polymers themselves are excluded, we can thus still assess whether large molecular mass polymeric products have to be expected through estimating the likeliness of polymerization for the most important products.
In the last step the results are summarized and a suggestion of the main components of the SEI is made based on the evaluation in the previous steps (2c).
Details on each step are given in the supplementary information and the suitability of the chosen estimators for kinetic factors is discussed in the following.

The approach is based on several assumptions. The reactions are assumed to be independent of the electrode. This, at the first glance, seems like a significant restriction. 
However, experimental studies showed that the surface chemistry on the graphene anodes in lithium-ion batteries does not differ considerably for nonactive electrodes or pure lithium electrodes. \cite{Xu2014} 
The transfer of the electrons is, therefore, not studied explicitly, but is just incorporated as excess electrons following in the footsteps of higher-level static studies. \cite{dft_electrolytes_tasaki-pre_2001_1,Korth2015} 
The transfer of electrons is modeled by the conservation of the modified charge in the overall reaction. 
As the mechanism is not the objective here, the approximation to take just the electron transfer into account seems reasonable.
The full problem is more complex, as for instance solvation details\cite{Tasaki2009} and the inhomogenity and dynamical nature of the SEI\cite{Lu2014b} are not taken into account.
The reactions constructed by our approach correspond to a macroscopic abstraction. All underlying microscopic steps are summarized in a simplified way into one reaction. While the mechanisms are still a vividly debated topic, the chemical composition is less uncertain. Though research is ongoing, the main components are well accepted and known since several years. The pioneer work by Aurbach and co-workers is still seen as valid, though several new species have been identified and the origin of some species has been put into question. \cite{Xu2014} The consent on the basic composition of the SEI offers a way to evaluate the performance of the proposed method. 
The redox fingerprint and the probability estimates are discussed in detail on the example of the reduction of ethylene carbonate (EC). 
EC is an indispensable solvent in state-of-the-art electrolytes for lithium-ion batteries, mainly because it forms a stable SEI. \cite{Xu2014} The composition of the SEI originating from EC was intensively studied and offers many references, but also sometimes contradictory results.
The complete redox fingerprint (input: maximal 2 reactant molecules, transfer of 1/2 electrons respectively) can be seen in table 3 in the supplementary information. For the overall evaluation of the redox fingerprint only the thermodynamically most favorable reactions are considered. By default, we consider reactions until half of the lowest reaction energy. For EC, three reactions are left on the electrolyte side (transfer of one electron) and eight reactions on the electrode side (transfer of two electrons). These reactions can be seen in table 1 sorted by energy (calculated with PM6-DH+). The reactions energies (as well as the oxidative stability and the solubility of individual compounds) can be calculated at different levels of theory. For an inclusion in a high-throughput approach the computational time is a critical factor. In our evaluation, we found a very high Pearson correlation ($>$ 0.99) of SQM (PM6-DH+) and hybrid DFT (B3-LYP/TZVP) calculations. Our approach offers the possibility to lithiate the compounds automatically until saturation (charge neutrality) to mimic the state in the SEI more closely, but a very high correlation ($>>$ 0.9) with the values of the unlithiated compounds offers the possibility to neglect lithiation.\\

\begin{scriptsize}
\begin{longtable}{p{4cm}p{1.5cm}p{1.5cm}p{2.0cm}p{2.0cm}p{2.0cm}}
\caption{Analysis of the 'redox fingerprint' of ethylene carbonate.} \label{tab:ECoverviewRFPA}\\
\hline
Reaction & Energy [kcal/mol] & Lowest HOMO [eV] & Tanimoto coefficient  & Minimal Number of Fragmentations & Minimal Number of Recombinations \\
\hline
\endhead
\hline
\endfoot
\multicolumn{6}{c}{Transfer of one electron} \\
\hline
1) \includegraphics[scale=0.15]{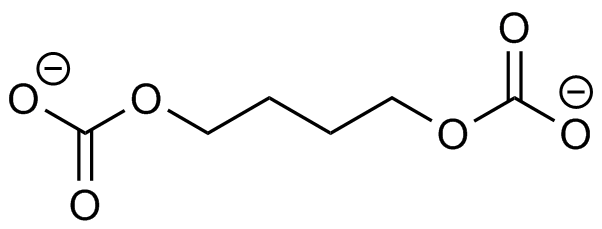} & -71.79 & -11.09 & 0.41  & 2 & 1  \\  
2) \includegraphics[scale=0.15]{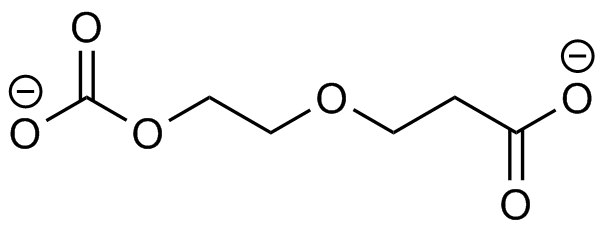} & -57.34 & -10.42 & 0.41 & 3 & 2 \\  
3) \includegraphics[scale=0.15]{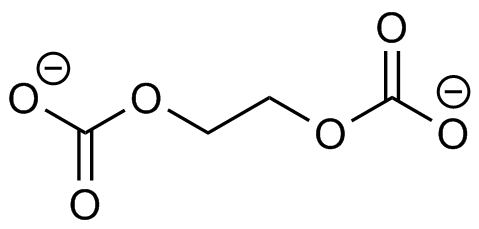}\ \ + \ \ \includegraphics[scale=0.15]{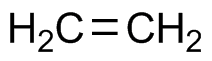} & -43.97 & -11.30 & 0.71 & 3 & 1 \\
\hline  
\multicolumn{6}{c}{Transfer of two electrons} \\
\hline
1)  \includegraphics[scale=0.15]{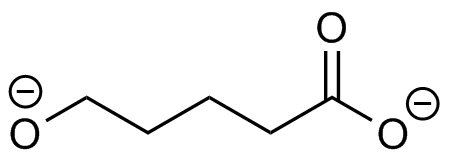}\ \ + \ \ \includegraphics[scale=0.15]{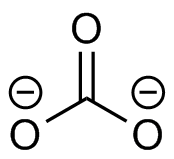} & -126.10 & -10.95 & 0.24 & 4 & 2 \\  
2)  \includegraphics[scale=0.15]{pic/org011110Y1.png} \ \ + \ \ \includegraphics[scale=0.15]{pic/org011110Y1.png} \ \ + \ \ \includegraphics[scale=0.15]{pic/org100001Y1.png}\ \ + \ \ \includegraphics[scale=0.15]{pic/org100001Y1.png} & -111.74 & -10.95  & 0.24 & 4 & 0 \\  
3) \includegraphics[scale=0.15]{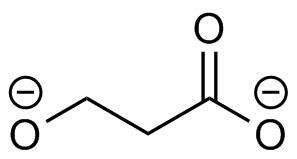}\ \ + \ \ \includegraphics[scale=0.15]{pic/org100001Y1.png}\ \ + \ \ \includegraphics[scale=0.15]{pic/org011110Y1.png} & -102.58 & -10.95 & 0.24 & 4& 1 \\  
4) \includegraphics[scale=0.15]{pic/rec01001000Y2.png}\ \ + \ \ \includegraphics[scale=0.15]{pic/rec01001000Y2.png} & -93.42 & -9.25 & 0.22 & 4 & 2 \\   
5) \includegraphics[scale=0.15]{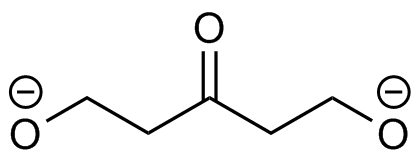}\ \ + \ \ \includegraphics[scale=0.15]{pic/org011110Y1.png} & -93.16 & -10.95 & 0.24 & 6 & 2 \\    
6) \includegraphics[scale=0.15]{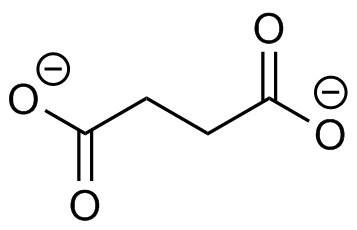}\ \ + \ \ \includegraphics[scale=0.15]{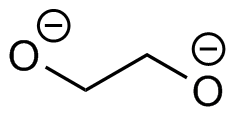} & -89.43 &-10.41 & 0.24 & 4 & 1 \\  
7) \includegraphics[scale=0.15]{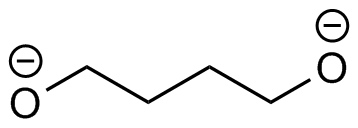}\ \ + \ \ \includegraphics[scale=0.15]{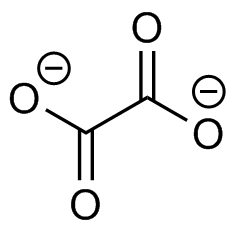}& -84.62 &-9.82 & 0.30 & 4 & 1 \\  
8) \includegraphics[scale=0.15]{pic/rec00002000Y3.png}\ \ + \ \ \includegraphics[scale=0.15]{pic/org011110Y1.png}\ \ + \ \ \includegraphics[scale=0.15]{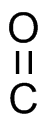} & -64.06 & -12.02 & 0.24 & 5 & 2 \\  
\end{longtable}
\end{scriptsize}

\bigskip

\noindent {\bf \fontfamily{phv}\selectfont{Results: SEI composition}}\\

\noindent
The redox fingerprint features all recently as important identified compounds. Examples are butylene dicarbonate (BDC, 1e/reaction 1), ethylene dicarbonate (EDC, 1e/reaction 3), carbonate ( 2e/reactions 1,2,3,5,8), ethylene (1e/reaction 3, 2e/reactions 2,3), succinate (2e/reaction 6) and oxalate (2e/reaction 7). 
Trace products include carbon monoxide and carbon dioxide, \cite{review_sei_novak_2010, Xu2014} which are present in the complete redox fingerprint. The presence of all important compounds is a first success for our approach. 
The applied heuristic rules are reflected in the output. The rules prohibit the formation of the recently reported orthoester compounds \cite{review_all_aurbach_2011b} and oligomerization. The importance of the orthoester compounds has not been revealed yet, but we assume no significant impact on the SEI features. The inclusion of oligo- and polymers seems more relevant, however. Some researchers attribute the stability of the SEI to its ability to form a branched polymeric network. \cite{dft_electrolytes_abraham_2013_1,dft_electrolytes_abraham_2013_2} 
A lot of possible, but not detected products are considered. This information is nevertheless not redundant, as only by taking it into account we can arrive at the correct solution in a black-box manner.
Neglecting it would only be possible through the inclusion of highly empirical assumptions, which would make the approach less suited for the investigation of unknown materials.

In comparison to the experimental results several reactions are combinatorially possible and would be thermodynamically allowed, but the products are not encountered in the SEI. The ranking of the reaction according to their thermodynamics is, therefore, not sufficient. 
We propose four estimators for the probability of a product to appear in the SEI that are readily available. Hereby, we circumvent high-level computations of the kinetics of the reactions. These computations would, on the one hand, be computationally costly and not feasible for a high throughput number. On the other hand, the redox fingerprint corresponds to a macroscopic abstraction and the identification of the reaction mechanism is non-trivial. The mechanisms leading to the most widely reported components of the SEI remain under debate after several years of investigation. \cite{aimd_sei_leung_cathode-material_2012,wft_electrolytes_leung_2013,Korth2015}
The thermodynamical basics are thus assessed at the level of quantum mechanics, while a heuristic approach is followed for estimating kinetic effects.

Our four estimators include the oxidative stability, the similarity to the reactant and the minimal number of required fragmentations and recombinations. 
Furthermore, we investigated the solubility of the compounds as an estimator in the spirit of Balbuena and co-workers. \cite{dft_electrolytes_tasaki-pre_2001_1} 
However, our calculations suggest that the solubility of all charged compounds in the polar solvent is very high and the difference in the amount of BDC and EDC can unlikely be attributed to it. 
(See supplementary information for a detailed discussion.)
Table 1 comprises the estimators for each reaction of EC. For reactions with more than one product, the most favorable estimator is chosen. The supplementary information features the results for all reaction products.
A high oxidative stability indicates that the back reaction is more unlikely than for compounds with a low oxidative stability. 
Korth showed recently that the orbital approximation is sufficient for a ranking purpose of the oxidative stability at SQM level. \cite{Korth2014} (B3LYP/TZVP and the values of the lithiated compounds can be found in the supplementary information.)
In this step only the ranking is relevant to identify oxidatively unstable compounds. The oxidative stability can therefore be approximated as the energy of the highest occupied molecular orbital (HOMO). 
For the one electron reactions, no significant difference in the oxidative stabilities can be found. EDC and BDC feature roughly the same oxidative stability. 
The (experimentally unobserved) ether compound has a slightly lower HOMO energy. For the two electron reaction, the experimentally less observed alkoxides have a poorer oxidative stability. 
The often observed oxalates and succinates as well as carbon monoxide, carbonate and ethylene are more stable. This allows the narrowing down of the two electron reactions, 
but it does not help distinguishing the relative amounts of BDC and (the experimentally mostly observed) EDC. 
Neither the solubility nor the oxidative stability is a sufficient probability estimator. The next estimators try to take kinetic features approximately into account. 
A first estimate is the degree to which the compounds have to transform in a reaction. One has to note that it is not claimed that complicated reaction mechanisms can be simplified into such an estimator. 
However, highly unlikely formations may be excluded in this way. An example is reaction 2 on the one electron side. The product formation would require the rotation of one of its fragments, 
which is highly unlikely to happen kinetically, but is possible combinatorially.  We propose the Tanimoto coefficient (also known as Jaccard index) as an estimator for the kinetics of a reaction. A similarity of the product to the reactant molecule can be taken as a first indication that no unlikely transitions are required.
Similar estimators are the number of the minimal required fragmentations and recombinations. The less bonds have to be broken per molecule, the less energy should be needed to overcome the activation barrier though the bond strength is currently not taken into consideration. Complicated structures mostly arise from options with a high number of fragmentations and recombinations. We do not claim that multi-step reactions are per se energetically unfavorable, but that they are more likely to be kinetically hindered than similar 'simple' reactions. 
The often experimentally identified compounds in the SEI formed from EC feature comparably high Tanimoto coefficients, especially EDC. Compounds that are not featured in the SEI as well as by-products as CO and ethylene have a low Tanimoto coefficient. That can be taken as an indicator that though the approximation is very rough, the Tanimoto coefficient can serve as an estimator to sort out unlikely reactions. 
The encountered products for the one electron transfer, EDC and BDC, are the  most favorable in terms of these kinetic estimates. 
The problem of explaining why EDC and not BDC dominates can thus not be solved with our approach, which is nevertheless in good agreement with high-level theoretical studies.

For the two electron transfer, the most unlikely products (and by-products) have a low Tanimoto coefficient, a high number of required fragmentations and recombinations, which can be seen as a success for the prediction. The main product, carbonate, is featured in many reactions and in many reactions that have a high Tanimoto coefficient and require a low number of transformations.
All results are then brought together in the final suggestions of the composition of the SEI as illustrated in table 2. To predict the polymerizability of a substance, different mechanisms that could lead to a polymerization would have to be studied, which is too complex to integrate in this approach. A simple estimator, whether a compound may be polymerizable or not, is included. If the input molecule incorporates double (triple) bonds, more than one functional groups with a double bond or a ring in which a functional group is incorporated, it is possibly polymerizable. The same applies for the species in the redox fingerprint. These criteria have been deduced from 'usual' polymerizable species. If a species is possibly polymerizable, it is indicated by a tick (\checkmark) behind the compounds in the final tables. In comparison with the experiment, EC features all relevant compounds, but those that are excluded by the constraints. On the electrolyte side, the RFPA suggests BDC, EDC and ethylene as the major products. All of these are observed though the outbalance of BDC in favor of EDC cannot be explained doubt free with our model, see below for a more detailed discussion. On the electrode side carbonate, ethylene, an alkoxide and oxalate are predicted. All of these products can be encountered in the SEI experimentally, but the amount of carbonate outweighs the other products. Polycarbonates cannot be observed due to the constraints, but our very simple polymerizability estimator suggests that polymerizability is possible. Several products like ethylene or oxalate are predicted as major products, while they are only observed minorly. This can be attributed to the requirement of atom mapping. As only primary reactions are investigated in our approach, the compounds that are necessary for atom mapping such as ethylene are featured often. However, these compounds might undergo subsequent reactions that are not explored here and therefore are only observed to a minor degree.
Taking only primary reactions into account could pose a serious limiation of our approach, but no problems are observed for the test cases investigated so far.
To further validate our approach, we examined several other standard electrolyte solvents (propylene carbonate PC, dimethyl carbonate DMC, dimethoxy methane DMM, 1,3 dioxolane 1,3-DL, tetrahydro furan THF). The results can be seen in table 2.\\

\begin{longtable}{p{2.75cm}p{2.75cm}p{2.75cm}|p{2.75cm}p{2.75cm}}
\caption{Suggestion for the composition of the SEI formed from standard electrolyte solvents on the basis of a redox fingerprint analysis.} \label{tab:StandardElectrolytesFin}\\
\hline
\multicolumn{3}{c}{RFPA estimated} & \multicolumn{2}{c}{Experimentally observed} \\
Electrode Side & Electrolyte Side & \multicolumn{1}{c}{Minor} & \multicolumn{1}{c}{Major} & Minor \\
\hline
\endhead
\hline
\endfoot
\multicolumn{3}{l}{EC \tinycheckmark}&&\\
\hline
\includegraphics[scale=0.15]{pic/org011110Y1.png} & \includegraphics[scale=0.15]{pic/rec00000020Y1.png} \tinycheckmark & $RCO_2^-$, CO & $CO_3^{2-}$, EDC & Polycarbonates, Oxalate, Succinate, Ethylene, Carbon Monoxide, Carbon Dioxide \\ \includegraphics[scale=0.15]{pic/org100001Y1.png} \tinycheckmark &\includegraphics[scale=0.15]{pic/rec00100010Y1.png} \tinycheckmark &&&\\

\includegraphics[scale=0.15]{pic/rec00002000Y3.png}&\includegraphics[scale=0.15]{pic/org100001Y1.png} \tinycheckmark &&&\\
\includegraphics[scale=0.15]{pic/rec02000000Y1.png} \tinycheckmark &&&&\\
\hline
\multicolumn{3}{l}{PC \tinycheckmark}&&\\
\hline
\includegraphics[scale=0.15]{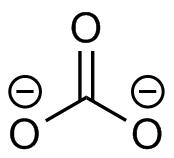}, 
\includegraphics[scale=0.15]{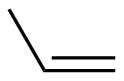}~\tinycheckmark , 
\includegraphics[scale=0.15]{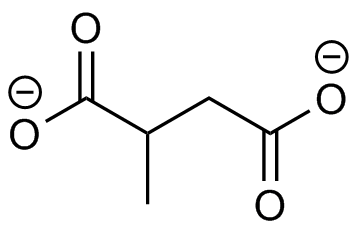}~\tinycheckmark , 
\includegraphics[scale=0.15]{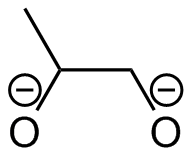} & 
\includegraphics[scale=0.15]{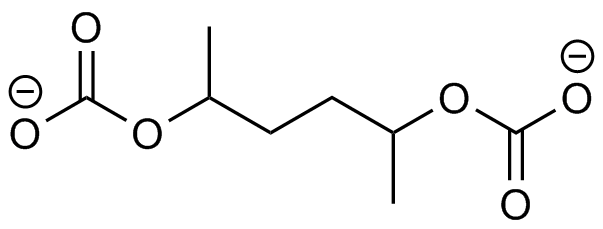}~\tinycheckmark , 
\includegraphics[scale=0.15]{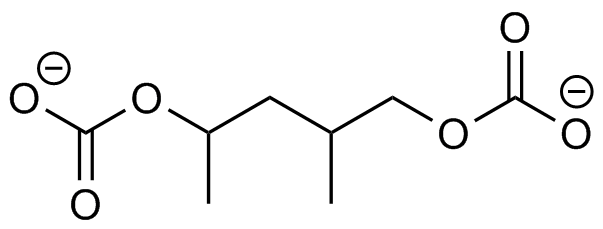}~\tinycheckmark , 
\includegraphics[scale=0.15]{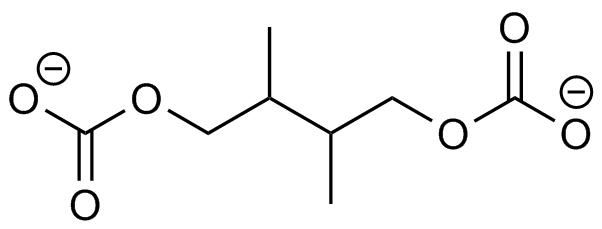}~\tinycheckmark , 
\includegraphics[scale=0.15]{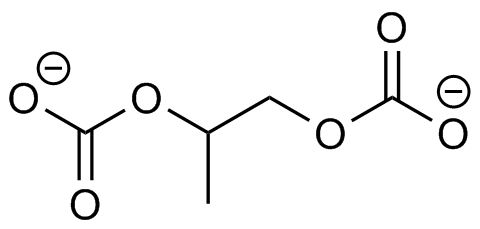}~\tinycheckmark , 
\includegraphics[scale=0.15]{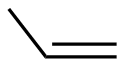}~\tinycheckmark
&  oxalate, alkoxides, $RCO_{2} ^{-} $, CO & $CO_3^{2-}$, HDC, PDC & $ROCO_2^-$, alkoxides, oxalate \\
\hline
\multicolumn{3}{l}{DMC}&&\\
\hline
\includegraphics[scale=0.15]{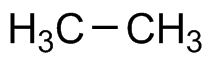}, 
\includegraphics[scale=0.15]{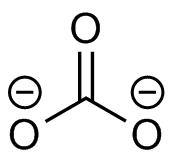}, 
\includegraphics[scale=0.15]{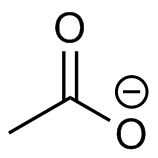},
\includegraphics[scale=0.15]{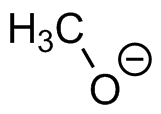}
& \includegraphics[scale=0.15] {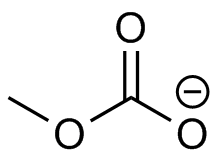}, 
\includegraphics[scale=0.15]{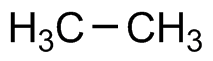}, 
\includegraphics[scale=0.15]{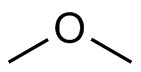}, 
\includegraphics[scale=0.15]{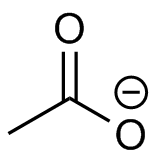}
& oxalate, CO & $CO_3^{2-}$, $MeOCO_2^-$, oxalate & dimethyl ether, alkoxides\\
\hline
\multicolumn{3}{l}{DMM}&&\\
\hline
\includegraphics[scale=0.15]{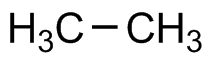}, 
\includegraphics[scale=0.15]{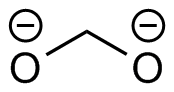},
\includegraphics[scale=0.15]{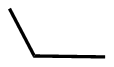},
\includegraphics[scale=0.15]{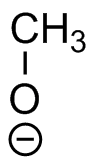}
& \includegraphics[scale=0.15]{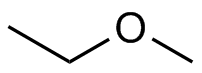},  
\includegraphics[scale=0.15]{pic/dmm1erec010010Y1.png},  
\includegraphics[scale=0.15]{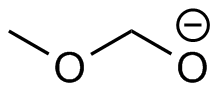} 
& dimethyl ether, ethan, alkoxides & alkoxides & \\
\hline
\multicolumn{3}{l}{1,3-DL \tinycheckmark}&&\\
\hline
\includegraphics[scale=0.15]{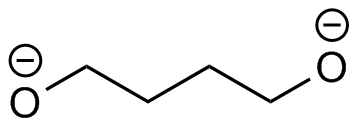},
\includegraphics[scale=0.15]{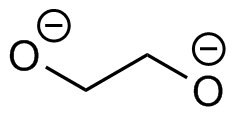}, 
\includegraphics[scale=0.15]{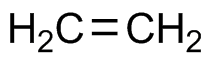}~\tinycheckmark ,
\includegraphics[scale=0.15]{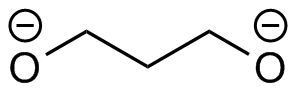},
\includegraphics[scale=0.15]{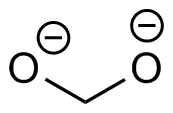}
& \includegraphics[scale=0.15]{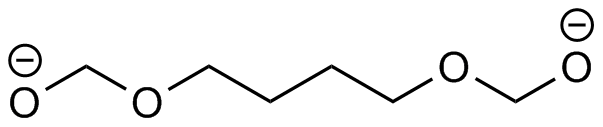}, 
\includegraphics[scale=0.15]{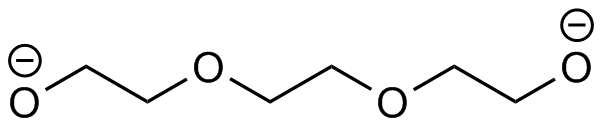},
\includegraphics[scale=0.15]{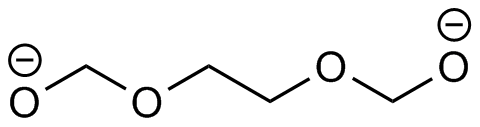},
\includegraphics[scale=0.15]{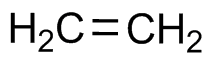}~\tinycheckmark ,
\includegraphics[scale=0.15]{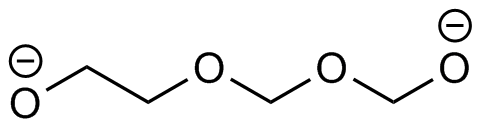}
& various alkoxides & alkoxides & \\
 \hline
\multicolumn{3}{l}{THF \tinycheckmark}&&\\
\hline
- & \includegraphics[scale=0.15]{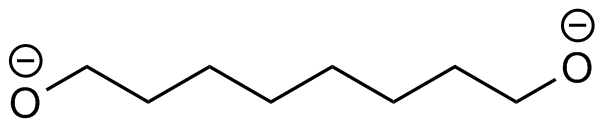} & - & alkoxides & \\
\end{longtable}

\bigskip

\noindent The thermodynamically favorable reactions for EC and PC are nearly identical, but more combinatorial possibilities emerge with the introduction of the methyl group. The formation of propylene dicarbonate (PDC) is predicted to be the likeliest reaction, which is the equivalent of ethylene dicarbonate in the EC RFPA. The equivalent of butylene carbonate is abbreviated 'HDC' (hexyl dicarbonate). No experimental literature data exists for pure PC on graphite anodes because it leads to the exfoliation. The data taken for comparison is for the pure lithium anode. The analogues of the EC species have been identified as the main components. \cite{Xu2014} On the electrolyte side, three derivatives of HDC as well as PDC and propylene are predicted. All of these compounds have been identified in experiments. Carbonate and proylene as well as a carboxylate and alkoxide are suggested for the electrode side. Carbonate is the experimentally most often observed compound out of these. 
The RFPA for DMC, a representative of the linear carbonates (other examples would be diethyl carbonate (DEC) or ethyl methyl carbonate (EMC)), exhibits all products that have been observed in the literature. 
For the one electron reduction the first reaction is assumed to take place preferably. This leads to the experimentally identified main product methyl carbonate. In the two electron reduction, carbonate is predicted as a main component along with ethane. Other experimentally found components are reproduced as well, like methanolate, dimethyl ether and oxalate. \cite{Xu2014}
Carbonate solvents are the most widely studied compound class, for other standard electrolyte solvents experimental data, especially of the pure compounds, is sparse. The main products are believed to be alkoxides for ether solvents. \cite{Xu2014} The ether solvents do not form a stable SEI. The RFPA of all ether solvents predicts various alkoxides as the main components along with ethers and alkanes/alkenes.
\\ \\ 

\noindent {\bf \fontfamily{phv}\selectfont{Estimating critical complex properties: Graphite exfoliation}}\\

\noindent
Overall, the RFPA is able to capture all major products observed in literature. However, the fingerprint itself is not enough to distinguish the different SEI formation properties of PC and EC. The fingerprint predicts the same structural features that also are predicted to have a similar probability to polymerize. This is not surprising because the electrode is not taken into account. PC forms a stable SEI on pure lithium, \cite{Xu2014} but leads to exfoliation of graphitic electrodes. The reason therefore has to be directly linked to features of the graphitic electrode. 
According to Shkrob {\it et al.} the differences in the formed polymers may be responsible for the different SEI retention capabilities. Tertiary radicals can be formed in propylene carbonate. They have a longer lifetime than the secondary radicals formed from EC and are more likely to recombine than to polymerize into extended networks. \cite{dft_electrolytes_abraham_2013_1,dft_electrolytes_abraham_2013_2}
These prediction have been made {\it ex situ} and were induced radiolytically. If this concept is also valid for the formation of SEIs on graphitic electrodes, the approach is going to be extended in this direction.
Another explanation for the EC/PC disparity was introduced by Besenhard and Winter. They suggested that this is caused not by the redox, but by the solvation properties of the two solvents: Co-intercalation of solvent molecules leads to graphite intercalation compounds (GICs) which are geometrically and/or energetically more favorable for keeping the anode intact in the case of EC. \cite{wft_electrolytes_tasaki-winter_2011,screening_dft_intercalation_tasaki_2014}
Subsequently, Han and co-workers have suggested the lithium-cation binding energy of (multiples of) solvent molecules as an estimator for the functionality of SEI films \cite{screening_dft_sei_han_2011} and Winter and Tasaki have suggested to investigate GICs for the same purpose. \cite{wft_electrolytes_tasaki-winter_2011,screening_dft_intercalation_tasaki_2014} We propose an estimator in the spirit of Winter and Tasaki, but now based on the results of the RFPA. 
The SEI is assumed to bestride the boundaries between the graphitic electrode and the electrolyte. 
We believe that the intactness of the electrode can be, at least partly, attributed to the stability/interlayer distance of the graphite intercalation compounds formed from the main redox products. We accordingly suggest a separate model system approach for estimating GIC properties in the following. Different model systems have been tested to simulate the GIC ranging from a coronen sandwich to a C150 double sandwich (See Supplementary Information for details) and concluded that a double coronen sandwich is best fitted for the prediction of the GIC stability. We reach a reasonable agreement with the results of Tasaki and Winter with this approach, calculating a double coronen sandwich with the molecule in question placed in the middle using PM6-DH+.
Table 2 shows the interlayer distance and the heat of formation of the double coronen sandwich complex for the major predicted compounds of the SEI via the RFPA as well as the pure solvents for EC and PC. (The GIC analysis for all products from EC and PC and a comparison of PM6-DH+ with DFT data can be found in the supplementary information.)
\\ \\

\begin{longtable}{ccc}
\caption{Proposed approach to estimate the graphite intercalation compound stability of RFPA predicted compounds via the interlayer distance of a double coronen sandwich and its heat of formation.} \\
\hline
Compound & Interlayer Distance (Maximum/Average) [\AA]& Heat of Formation [kcal/mol] \\
\hline
\endhead
\hline
\endfoot
\multicolumn{3}{c}{Ethylene Carbonate}\\
\hline
EC/$Li^+$* & 5.9 & -11.23 \\
EC/$Li^+$ & 6.41/6.28 & -9.29 \\
\includegraphics[scale=0.15]{pic/rec00000020Y1.png} & 7.51/7.34 & -11.18 \\ 
\includegraphics[scale=0.15]{pic/rec00100010Y1.png} & 7.21/6.83 & -12.18 \\ 
\includegraphics[scale=0.15]{pic/rec00002000Y3.png} & 6.94/6.92 & -2.32 \\ 
\includegraphics[scale=0.15]{pic/rec02000000Y1.png} & 6.69/6.54 & 15.86 \\ 
\includegraphics[scale=0.15]{pic/org100001Y1.png} & 6.24/6.22 & -4.01 \\ 
\includegraphics[scale=0.15]{pic/org011110Y1.png} &6.03/5.77 & 10.35 \\ 
\hline
\multicolumn{3}{c}{Propylene Carbonate}\\
\hline
PC/$Li^+$* & 7.0 & -5.72 \\
PC/$Li^+$ & 7.41/7.11 & -4.40 \\
\includegraphics[scale=0.15]{pic/pc1erec00000001001Y1.png} & 8.59/8.44 & -15.13\\ 
\includegraphics[scale=0.15]{pic/pc2erec02010000000Y9.png} & 8.39/8.12 & -5.75\\ 
\includegraphics[scale=0.15]{pic/pc1erec00000002000Y1.png} & 8.25/8.02 & -13.83\\ 
\includegraphics[scale=0.15]{pic/pc1erec00000000002Y1.png} & 8.22/8.10 & 5.41\\ 
\includegraphics[scale=0.15]{pic/pc1erec00100000001Y1.png} & 7.46/7.32 & -0.86\\ 
\includegraphics[scale=0.15]{pic/pc2eorg1110011Y1.png} & 7.09/6.97 & 6.08\\ 
\includegraphics[scale=0.15]{pic/pc2eorg1100001Y1.png} & 6.75/6.54 & -6.52\\ 
\includegraphics[scale=0.15]{pic/pc2eorg0011110Y1.png} & 6.03/5.77 & 10.35 \\ 
\hline
\multicolumn{3}{l}{* obtained from reference \cite{wft_electrolytes_tasaki-winter_2011}}
\end{longtable}

\noindent {\bf \fontfamily{phv}\selectfont{Results: Graphite exfoliation}}\\

\noindent
Perusing table 2 the following observations can be made: Comparing the solvent/lithium complexes from our model system setup with the periodic boundary condition DFT calculations of Tasaki and Winter,
we predict slightly (about half an Angstr\"om) higher interlayer distances (ILDs) and somewhat (1.5 to 2 kcal/mol) lower heats of formation (HOFs), but very similar relative differences between EC and PC.
In general, the interlayer distances for the reduction products of EC are (about one Angstr\"om) smaller than those of PC, which is supposed to lead to more stable GICs.
The most unfavorable compound in terms of the interlayer distance is in both cases the experimentally less observed BDC and its analogon HDC.
Three different isomers of HDC exist, two in which the methyl groups are in 1,4 and 2,3 trans configuration, one in which they are in 1,3 cis configuration.
The ILD the most unfavorable (cis-1,3) HDC is on average 1.1 Angstr\"om bigger than for BDC, the ILD of PDC is on average only 0.5 Angstr\"om bigger than for EDC.
The different HDC isomers show rather similar ILDs, but the trans-1,4 configuration is not giving stable GICs.
Very interesting are the findings, that (2,3-trans and 1,3-cis) HDC forms more stable GICs than BDC despite the somewhat higher ILD,
and that while BDC and EDC form similar stable GICs, it is not the same as for HDC and PDC, as the PDC-GIC is barely stable.
This difference likely contributes to the preferred formation of the (also thermodynamically more likely) HDC,
but does not explain the preferred formation of the (thermodynamically less likely) EDC instead of BDC.
Our suggestion for an estimator that can be applied in large scale screening approaches is to evaluate the ILD distances of the GIC with the most favorable HOFs
from all those compounds that were predicted as SEI components by the RFPA analysis.
In the case of EC EDC would have to be considered, with a maximum ILD of 7.21 Angstr\"om, for PC only 1,3-cis HDC would have to be considered with a maximum ILD of 8.59 Angstr\"om.
Dispersion forces, which are the main 'glue' between graphite sheets are decaying according to $R^{-6}$,
resulting in large contributions in the region of 3 to 5 Angstr\"om, smaller ones up to about 7 Angstr\"om, but smaller effects at larger distances.\cite{DISP}
The ILD differences between EC/Lithium (5.9 Angstr\"om) and PC/Lithium (7.0 Angstr\"om) are therefore much less significant
than what we observe the EDC-GIC (7.21 Angstr\"om) and the HDC-GIC (8.59 Angstr\"om),
which is why we believe that our approach should be much better suited for estimating graphite exfoiliation effects when screening electrolyte materials.
\\ \\

\newpage
\noindent {\bf \fontfamily{phv}\selectfont{Conclusions}}\\

\noindent We presented a possibility to tackle the prediction of complex battery electrolyte features like solid-electrolyte interphase (SEI) formation and graphite exfoliation systematically and automatically.
Our redox fingerprint analysis (RFPA) is based on the generation of all possible reactions,
applying combinatorial considerations, heuristic rules and specified constraints concerning the reactant number and the number of transferred electrons.
The reactions are subsequently analyzed, reaction energies are determined and
the probability to encounter the products of the most favorable reactions in the SEI is predicted through their solubility, oxidative stability and estimators for their kinetics. 
We introduced the Tanimoto coefficient and the required number of transformations as possibilities to rule out kinetically hindered reactions.  
The comparison of the outcome of the RFPA with experimental results shows that, when experimental results are present, they are reproduced well.
We furthermore applied the stability of automatically generated graphite intercalation compounds (GICs) of all reactants and relevant products as an estimator for graphite exfoliation.
Standard electrolyte compounds are discussed in detail to show that the predictions of our approach agree with experimental findings and theoretical studies at higher level.

Our approach is well suited for the inclusion in screening strategies to identify new electrolyte solvents for the application in batteries.
Beyond a pure screening approach these two features are still highly investigated, theoretically and experimentally, but until now not systematically, bestriding compound classes.
The presented approach offers the opportunity to estimate the major components of a SEI for an arbitrary molecule or mixture of molecules and predict the likeliness of them solvent
to lead to graphite exfoliation based on the stability of the graphite intercalation compounds formed from these components.
In the future, the stability of the SEI may now be correlated to its predicted composition, which is only possible in combination with systematic experimental investigations.
Apart from substantially broadening the possibilities of electrolyte screening, this approach is also perfectly fitted to aid in-depth experimental and theoretical investigations:
Our approach offers an easy way to systematically identify all likely compounds and promising reactions beforehand
which is helpful for both the set-up and the analysis of experiments,
as well as the preparation of static {\it ab initio} studies, for which reactions are currently handpicked based on chemical intuition.
An extension of our ansatz to oxidative decomposition reactions on the cathode side is in preparation.
Combinatorial quantum chemistry thus significantly broadens the horizon of computational screening for molecular organic electrolyte materials
and offers the possibility to aid systematic experimental and theoretical investigations.
\\ \\

\noindent {\bf \fontfamily{phv}\selectfont{Acknowledgment.}}\\

\noindent The authors would like to thank Gerhard Maas for helpful discussions. Financial support from the Barbara Mez-Starck Foundation is gratefully acknowledged.
\\ \\

\noindent {\bf \fontfamily{phv}\selectfont{Supporting information}}\\

\noindent Full RFPA results, solubilities and oxidative stabilities for all compounds are available for download from ...
\\ \\

\bibliographystyle{achemsolx2}
\fontfamily{phv}\selectfont{\bibliography{korth_sei}}

}
\end{document}